\documentstyle[aps,prl,epsf,preprint]{revtex}
\def\la{\mathrel{\mathpalette\fun <}}
\def\ga{\mathrel{\mathpalette\fun >}}
\def\fun#1#2{\lower3.6pt\vbox{\baselineskip0pt\lineskip.9pt
        \ialign{$\mathsurround=0pt#1\hfill##\hfil$\crcr#2\crcr\sim\crcr}}}

\begin{document}


\title{\vskip-2.5truecm{\hfill \baselineskip 14pt {{
\small  FERMILAB-Pub-95/400-A\\
       \hfill SISSA--163/95/EP \\
\hfill December 1995}}\vskip .1truecm}
 {\bf Vacuum Decay along} 
\vskip 0.1truecm {\bf Supersymmetric Flat Directions
}}


\author{Antonio Riotto\thanks{
 riotto@fnas01.fnal.gov}$^{(1)}$ and Esteban Roulet\thanks{
roulet@susy.he.sissa.it}$^{(2)}$}

\address{$^{(1)}${\it NASA/Fermilab Astrophysics Center, Fermilab 
National Accelerator Laboratory,\\Batavia, IL 60510, USA}}

\address{$^{(2)}${\it International School for Advanced Studies, SISSA-ISAS,\\
Via Beirut 2/4, I-34014, Miramare, Trieste, Italy}}

\maketitle


\begin{abstract}
\baselineskip 12pt
It has been recently realized  that  within the Minimal
Supersymmetric Standard Model, for certain  patterns of 
superpartner masses, consistent with all the present 
 experimental constraints, the scalar potential may develop  at some
scale $Q_0$ unbounded 
color/charge breaking directions involving the sfermion fields, and that these 
patterns are then excluded unless some new physics is invoked at or 
below the scale $Q_0$. We reanalyze this observation and point out
that  such patterns of superpartner 
masses at the weak scale are  {\it not} ruled out when taking 
into account the probability of decay for the metastable color 
conserving minimum
along these color breaking unbounded directions. It turns out that 
the color conserving minimum, although
metastable, has  a lifetime longer than the present age of the
Universe and can survive both quantum tunneling and the effects of high 
temperatures in the early Universe, causing the color/charge breaking effects
to be in practice not dangerous.

\end{abstract}

\thispagestyle{empty}

\newpage
\pagestyle{plain}
\setcounter{page}{1}
\def\beq{\begin{equation}}
\def\eeq{\end{equation}}
\def\beqa{\begin{eqnarray}}
\def\eeqa{\end{eqnarray}}
\def\tr{{\rm tr}}
\def\x{{\bf x}}
\def\p{{\bf p}}
\def\k{{\bf k}}
\def\z{{\bf z}}
\baselineskip 20pt
\begin{flushleft}
{\large\bf 1. Introduction}
\end{flushleft}
\vspace{0.5cm}
 
There are many good reasons to believe that the Standard Model is not 
the ultimate theory  of nature since it is unable to answer many 
fundamental questions. One of them, why and how the electroweak scale 
and the Planck scale are so hierarchically separated has motivated the
Minimal Supersymmetric extension of the Standard Model (MSSM) as the 
underlying theory at scales of order 1 TeV 
\cite{haber}.  A  huge number of new parameters appear when 
considering the  MSSM.  Some of these new
parameters are constrained by the unsuccessful searches of new
particles at accelerators. Others 
may receive severe bounds from the requirement  of avoiding large 
flavour-changing neutral currents. Moreover, constraints on the
parameter space, which mostly involve the soft supersymmetry breaking 
trilinear terms $A$'s,  arise from the existence of charge and/or
color breaking minima in the scalar sector  when looking  at some 
particular directions with nonvanishing vacuum expectation values 
(VEV's) of the Higgs fields \cite{color}.

By considering directions in the field space which do not interest 
the VEV's of the Higgs fields, 
it has been recently pointed out that certain mass patterns for the 
superpartners cannot arise at low energy unless there is new physics 
beyond the MSSM and below the grand unified (GUT) scale \cite{olive}. 
 Indeed,  there are many flat directions in the field space of the MSSM and 
it may happen that some combination of the squark and/or slepton 
mass-squared parameters 
get negative at some scale $Q_0$ below the GUT scale when running 
through the Renormalization Group Equations (RGE's) from the weak 
scale up. This leads either to the appearance of 
unacceptable color breaking minima or to unbounded
from below directions in the effective potential for the squark 
and/or slepton fields $\phi$'s, making the color conserving minimum 
$ \phi=0$ metastable.
 In such a case, it has been argued in ref.~\cite{olive} that 
the corresponding region of the parameter space is 
either ruled out or there must exist some new physics below the scale 
$Q_0$ whose effects can change the evolution of the mass-squared 
parameters with the scale $Q$ \cite{olive}.  

The situation here is fairly analogous to what happens for the
effective 
potential of the SM Higgs field $H$: for a top quark mass large 
compared to the Higgs and gauge bosons masses, the one--loop top quark
contribution to the effective potential will dominate the others 
and drive the
coefficient of the quartic term $H^4$ negative for very large values 
of $H$, thus destabilizing the effective potential  and making 
our vacuum  at $ \langle H\rangle\simeq 250$ GeV a local, but not
global,  minimum 
\cite{sher}. Nonetheless, the electroweak vacuum need 
not be absolutely stable. For certain top quark mass $M_t$ and Higgs 
boson mass $M_H$  it may just be  instead metastable, as long as its 
 lifetime exceeds the present age of the Universe \cite{meta}. 
The decay of the electroweak vacuum may be driven at low temperatures 
by quantum tunneling
or at high temperatures by thermal excitations. 
Even if the requirement that our vacuum survives the high temperatures 
of the early Universe places  strong
constraints from vacuum stability on the $(M_t,M_H)$ parameter space,
still values of the top quark and Higgs boson masses for which our 
vacuum is metastable, but with a lifetime larger than the present 
age of the Universe, are allowed. 

The purpose of the present Letter is to reanalyze  the constraints 
on supersymmetric models discussed in ref. \cite{olive}, 
involving some combination of the sparticle masses at the weak scale 
and imposed to avoid 
large color breaking VEV's  $ \phi\neq 0$ or destabilized effective 
potentials along some squark and/or slepton directions,  and to point out that 
such limits may be  weakened by considerations about the survival of the
color conserving minimum. Even if such a minimum may be  metastable 
in large regions of the parameter space  of the superpartner masses, 
its lifetime turns out to be almost everywhere longer than the present 
age of the Universe. This means that the regions of the parameter
space ruled out (unless some new physics appears before  a certain 
scale $Q_0$) in ref. \cite{olive}   may  be indeed permitted without 
any need of new physics.

The paper is organized as follows: In Section {\bf 2} we shall briefly
review the effective potential along some particular directions of 
the squark fields  
showing why it may be destabilized and which are the consequent constraints
on the superpartner mass patterns \cite{olive}. In Section {\bf 3} we 
discuss the color conserving $\phi=0$ decay rate at zero temperature,
leaving the finite temperature case to Section {\bf 4}. Finally, 
Section {\bf 5} presents our conclusions.

\vspace{0.5cm}
\begin{flushleft}
{\large\bf 2. The effective potential and flat directions}
\end{flushleft}
\vspace{0.5cm}

Let us consider the same flat direction in the squark fields analysed 
in ref. \cite{olive}: $\tilde{u}_R^r=\tilde{s}_R^g=\tilde{b}_R^b\equiv
\phi/\sqrt{2}$. Along this particular direction the coefficient 
$\lambda(Q)$ of the quartic term $\phi^4$ is vanishing for all scales 
$Q$, and the one--loop effective potential reads
\begin{equation}
\label{pot}
V(\phi)=\frac{1}{2}m^2(Q)\phi^2(Q)+V_{{\rm 1-loop}}(\phi),
\end{equation}
where $m^2\equiv m^2_{\tilde{u}_R}+m^2_{\tilde{s}_R}+m^2_{\tilde{b}_R}$ and 
$V_{{\rm 1-loop}}$ is the one--loop correction to the effective potential 
(in the $\overline{DR}$-scheme) 
\begin{equation}
V_{{\rm 1-loop}}(\phi)=\frac{1}{64\:\pi^2}\:{\rm Str}\:
{\cal M}^4(\phi)\left[{\rm ln}\left(\frac{{\cal M}^2(\phi)}{Q^2}\right)-\frac{3}{2}\right].
\end{equation}
where Str counts properly all the degrees of freedom,
 summing over all the mass eigenstates which get mass in 
the $\phi$-background field.

Since the one--loop potential behaves as
$\propto{\rm ln}\left(g_3^2\phi^2/Q^2\right)$ for large values of 
$\phi$ ($g_3$ is the $SU(3)$ gauge coupling constant), 
in order that the approximation 
of neglecting loop
effects be safe one should adopt in the renormalization group improved
tree level potential a scale $Q\simeq g_3\phi$ to make the logarithms
small. 
In practice, we will adopt
$Q=Q_\phi\equiv\sqrt{g_3^2\phi^2+M_3^2}$, so as to stop the running 
below $Q\simeq
M_3$, and use for the effective potential

\begin{equation}
\label{eff}
V(\phi)=\frac{1}{2}m^2(Q_\phi)\phi^2.
\end{equation}
The prescription is trivial: just evaluate the mass parameter $m^2$ at
the scale $Q\simeq Q_\phi$.

The RGE for $m^2$ is given by 
\begin{equation}
Q\frac{d m^2}{d Q}=\frac{1}{8\:\pi^2}\left[-16\:g_3^2\:M_3^2-
\frac{8}{3}\:g_1^2\:M_3^2+
2\:h_b^2\:\left(m^2_{\tilde{q}_L}+m^2_{\tilde{b}_R}+
m^2_{H_1}+A_b^2\right)\right],
\end{equation}
where $g_1$ is the standard $U(1)$ 
coupling, $M_i$ are gaugino masses, $h_b$ is the 
bottom quark Yukawa coupling and $A_b$ is the bottom quark 
trilinear mixing parameter, and all parameters are running. 
If $\tan\beta$ (the ratio of the two 
Higgs VEV's) is not too large,
$h_b$ is small and the term proportional to $h_b^2$ in the above 
Equation can be neglected, leading to the solution  \cite{olive}
\begin{eqnarray}
m^2(Q)&=&m^2-\frac{2}{\pi^2}g_3^2(M_3)M_3^2{\rm ln}(Q/M_3)
\left\{\frac{1+3 g_3^2(M_3){\rm ln}(Q/M_3)/(16\pi^2)}{
\left[1+3 g_3^2(M_3){\rm ln}(Q/M_3)/(8\pi^2)\right]^2}
\right\}\nonumber\\
&-&\frac{1}{3\pi^2}g_1^2(M_1)M_1^2{\rm ln}(Q/M_1)
\left\{\frac{1-11 g_1^2(M_1){\rm ln}(Q/M_1)/(16\pi^2)}{
\left[1-11 g_1^2(M_1){\rm ln}(Q/M_1)/(8\pi^2)\right]^2}\right\},
\label{mq.eq}
\end{eqnarray}
where all masses on the right-hand side are physical 
(propagator pole) masses and the only undertermined factors are 
$g_1^2(M_1)$ and $g_3^2(M_3)$, which depend on the full spectrum 
of the superpartner masses and the initial condition 
$\alpha_3(M_Z)=0.12$. If one takes $M_1\leq M_3$, as usually results
in GUT models, the effects of the term proportional to $g_1^2$ in
Eq. (\ref{mq.eq}) is negligible. To obtain $g_3(M_3)$ we have assumed a
common supersymmetric threshold at $M_3$, but the results are 
almost unsensitive to this simplifying assumption.

In Fig. 1 we show the contours of 
$m^2(Q_0)=0$ for different values of $Q_0$ in the 
$(m/\sqrt{3},M_3)$--plane. Portions of the parameter space lying 
to the right of each contour (for a given  scale $Q_0$) are 
characterized by an effective potential  unbounded from below and by
a metastable color conserving minimum  $ \phi=0$, unless some new 
physics capable of modifying this situation 
is present between the soft supersymetry breaking scale 
($\sim$ 1 TeV) and $Q_0$\footnote{The contours of Fig. 1 are 
slightly different from those presented in ref. \cite{olive} since 
there the authors have introduced a non-renormalizable operator 
$(1/6)\phi^6/M^2$ which lifts
the unbounded direction and results in a VEV  $\phi(Q)=M^{1/2}
(-m^2(Q))^{1/2}$. Then they have  plotted the self-consistency 
condition $\phi(Q)= Q$ and found that for $m/\sqrt{3}\la 0.7\: 
M_3$ some new physics below $M_{Pl}$ is necessary because of the existence of 
large color breaking VEV's.}. It is easy to see, from Eq. (\ref{mq.eq}),
that for $m\leq M_3$ one has that the value of the field $\phi_0$ for
which $V(\phi_0)=0$ is
\begin{equation}
\phi_0\simeq M_3\:{\rm exp}\left[ {\pi^2m^2\over 2g_3^2M_3^2}\right].
\label{phi0.eq}
\end{equation}

In Fig. 2 we draw the running scalar mass $m^2(Q_\phi)$, normalized to
its value $m^2$ for $\phi=0$, as a function of the field $\phi$,
assuming that $m/\sqrt{3}=500$ GeV and for values of $M_3$ of 1500,
1000, 700 and 600~GeV. The
effective potential is just obtained by multiplying the result by
$m^2\phi^2/2$. 
 Notice that a barrier will be present in $V(\phi)$ 
separating the metastable vacuum 
$\phi=0$ from the region of very large values of $\phi$, where
$m^2(Q_\phi)<0$ and  the potential becomes negative and unbounded
from below.

 From the analysis of the effective potential one can then 
conclude that the sparticle masses at the weak scale must 
satisfy severe relations among each other to avoid the color 
conserving minimum $ \phi=0$  becoming a local minimum and the 
potential to develop unbounded directions for large values of the 
field $\phi$. 

In the next Sections we shall show that these
relations may  be relaxed, since in most of the region in the
parameter space $(m/\sqrt{3},M_3)$ where the minimum
 $\phi=0$ is metastable, it has however a lifetime longer then the age
of the Universe, so that color breaking effects are in practice not
dangerous. 
\vspace{0.5cm}
\begin{flushleft}
{\large\bf 3.  Nucleation by quantum tunneling}
\end{flushleft}
\vspace{0.5cm}
The decay of the metastable color conserving minimum  $\phi=0$ may
occur by the nucleation of bubbles of the unstable phase. 
If the bubble is too small, it collapses under its surface tension. 
If the bubble is large enough, it expands classically, eventually 
absorbing all the metastable phase. At zero temperature, the vacuum 
can decay only by quantum tunneling through the barrier separating 
the metastable vacuum from the region of negative values of
$V(\phi)$. 
The WKB amplitude for false-vacuum decay by tunneling may be 
found expanding the Euclidean path integral about the {\it bounce} 
solution to the 
Euclidean equation of motion \cite{coleman}
\begin{equation}
\partial^2 \phi=\frac{d V(\phi)}{d\phi}.
\end{equation}
The bounce solution is an $O(4)$ rotationally symmetric solution and solves
\begin{equation}
\label{bounceeq}
\left(\partial_s^2+\frac{3}{s}\partial_s\right)\phi=\frac{d V(\phi)}{d\phi},
\end{equation}
where $s=(t_E^2+{\bf r}^2)^{1/2}$. It takes some convenient 
value $\phi(0)$ at $s=0$ (with $\phi^{\prime}(0)=0$), probing 
the unstable region of the potential ($\phi(0)>\phi_0$, where
$V(\phi_0)=0$), 
and falls to the false vacuum $\phi=0$ as $s\rightarrow\infty$. 
When viewed as a function of the Euclidean time $t_E$, 
the bounce solution interpolates between the false vacuum 
$\phi(t_E\rightarrow-\infty,{\bf r})=0$ and an  unstable
 bubble $\phi(t_E=0,{\bf r})$, which is just large enough to expand 
on its own classically. The Euclidean action $S_4$ of the bounce 
solution yields the exponential
suppression of the rate for the false vacuum decay per unit volume
\begin{equation}
\Gamma_4/V= D_4\:{\rm exp}\left[-S_4\right],
\label{gamma4.eq}
\end{equation}
where $D_4$ is a coefficient calculated from fluctuations around the bubble. 

The rate for the decay of the metastable color conserving minimum $\phi=0$
at zero temperature may be found simply by solving Equation (\ref{bounceeq}) 
numerically, using the effective potential (\ref{eff}) discussed
earlier, 
and computing the Euclidean action. Some qualitative 
flavour of the results can be
obtained by noting that, if $\phi_0\gg m$ and $M_3$, 
 the bounce is mainly determined by the
behaviour of $V$ at large values of $\phi$, near $\phi_0$. Since 
$\left.dV/d\phi\right|_{\phi_0}\propto M_3^2\phi_0$, 
we see from Eq.~(\ref{phi0.eq})
 that the radius of the bounce is $\sim M_3^{-1}$, and
for instance the kinetic energy contribution to $S_4$  results
$\sim \phi_0^2/M_3^2\simeq {\rm exp}\left(\pi^2m^2/(2
g_3^2M_3^2)\right)$, which is quite large unless $m\ll M_3$.

To decide whether the metastable vacuum $\phi=0$ would survive the 
age of the present Universe $\sim 10^{10}$ yr, one has to multiply 
Eq.~(\ref{gamma4.eq}) by the space-time volume
of the past light cone of the observable Universe. The  condition 
that the lifetime of the metastable state $\phi=0$ is longer than 
the present age of the Universe translates into (neglecting the very 
weak dependence upon the prefactor $D_4$)
\begin{equation}
S_4\ga 400. 
\end{equation}
Fig. 3 shows the contours  in the $(m/\sqrt{3},M_3)$--plane for 
different values of $\log_{10}(S_4)$, the critical value $S_4=400$ 
corresponding to the contour labelled by 2.6. Regions
to the right of each contour are characterized by an Euclidean 
action  smaller than  the value at the contour. A comparison 
between Figs. 1 and 3 shows that the color symmetric vacuum 
$\phi=0$, although  metastable,  
has  indeed a lifetime which exceeds  
 the present age of the Universe, in 
all the regions of the $(m/\sqrt{3},M_3)$--plane, with the exception 
of  a very small wedge at small values of $m$. From Fig.~1, one sees
that in this wedge the vacuum $\phi=0$ could become stable only by 
the introduction
of new physics at a scale below $\sim 10^4$~GeV. 

 From the analysis of the metastable vacuum decay by quantum tunneling 
at zero temperature we may therefore conclude that apparently
dangerous 
flat and unbounded directions in the squark $\phi$-field space are 
indeed safe: the barrier separating the metastable vacuum from the 
regions of negative values of $V(\phi)$ inhibits the
formation of sufficiently large bubbles and the  color conserving 
vacuum $\phi=0$  survives 
the quantum tunneling. 

In the next Section we shall analyze the fate of the color conserving  
minimum when considering the thermal effects present in the early Universe.
\vspace{0.5cm}
\begin{flushleft}
{\large\bf 4. Nucleation by thermal excitation}
\end{flushleft}
\vspace{0.5cm}

Another source of energy to cross the barrier is  the high temperature
of the early Universe. In the high-temperature plasma, thermal 
fluctuations may excite a bubble sufficiently large so as to be able
to induce the transition, and the
probability 
of a thermal fluctuation of associated energy $E_b$ to 
cross the barrier is simply given by a Maxwell-Boltzmann suppression 
factor ${\rm exp}(-E_b/T)$. One may think that the rate will not be 
exponentially 
suppressed at high temperatures, large compared to the barrier 
energy, but however the effective potential, and therefore $E_b$, 
also depends on the temperature making the barrier higher. At 
high temperatures, therefore, there is a thermal energy to cross 
the barrier, but the barrier is higher. 

We now need to find the energy barrier for the phase transition. 
For this we need a bubble corresponding to a static, unstable 
solution of the classical equation of motion \cite{linde}
\begin{equation}
\label{bounce}
\left(\partial_r^2+\frac{2}{r}\partial_r\right)\phi=\frac{d V(\phi,T)}{d\phi},
\end{equation}
where $V(\phi,T)=V(\phi)+\Delta V_T(\phi)$ is obtained by adding 
to the zero temperature effective 
potential the one--loop finite--temperature correction \cite{dj}
$\Delta V_T(\phi)\equiv V_T(\phi)-V_T(0)$, with
\begin{equation}
V_T(\phi)={T^4\over 2\pi^2}\sum_i\pm n_i\int_0^\infty dq\  q^2{\rm
ln}\left[ 1\mp{\rm exp}\left(-\sqrt{q^2+m_i^2(\phi)/T^2}\right)\right].
\end{equation}
Here the upper sign is valid for bosons and the lower one for fermions,
and $n_i$ are the corresponding degrees of freedom.
 Notice that
substracting $V_T(0)$ from the effective potential is 
necessary for computing the bounce action. 

It is easy to see that $V_T(0)=-\pi^2g_*T^4/90$, with $g_*=n_b+(7/8)n_F$
counting the degrees of freedom lighter than $T$. Hence, for $\phi\gg
T$,  $\Delta V_T(\phi)=\pi^2\bar g_*T^4/90$, where $\bar g_*$ counts the
particles heavier than $T$ in the presence of the field $\phi$ and
lighter than $T$ in $\phi=0$. In our case, $g_*=60$ for $T\gg
m$ and $ M_3$.
On the other hand,  one has for $T\gg \phi$,
\begin{equation}
\Delta V_T(\phi)\simeq {T^2\over 24}\sum\left[n_B\Delta m_B^2+{n_F\over
2}\Delta m_F^2\right],
\end{equation}
where $\Delta m^2\equiv m^2(\phi)-m^2(0)$, so that only the $\phi$
dependent mass terms contribute. In our case $\Delta V_T(\phi)\simeq 2
g_3^2 T^2\phi^2$ for $T\gg \phi$, getting its dominant contributions 
from the gluons (via seagull terms), the gluinos and the $u_R^r$,
$s_R^g$ and  
$b_R^b$ quarks  (from the $\tilde g q \tilde q$ couplings)
and the corresponding squarks  (via the $D$--terms), 
with the exception of the flat
direction\footnote{we have neglected the small
contribution from the photon and photino.}. 

Again, we have solved   Eq. (\ref{bounce}) numerically searching for 
a solution which probes
unstable values of $\phi$ at $r=0$ and falls off to the 
metastable vacuum $\phi=0$ as $r\rightarrow\infty$. 
The probability of tunneling per unit time per unit
volume is given by
\begin{equation}
\Gamma_3/V=D_3\:{\rm exp}\left[-S_3/T\right],
\end{equation}
where $D_3$ is the determinant factor and $S_3$ is the 
three-dimensional action of the solution of Eq. (\ref{bounce}). 
As discussed by Anderson in ref. \cite{meta} we should now multiply 
by the volume our current horizon had when at temperature $T$, which 
is $V(T)\sim (10^{10}\:{\rm yr})^3\times(3\:{\rm K}/T)^3$, and by
the amount of time the Universe spent at temperatures $T$, which 
is $t\sim M_{{\rm Pl}}/T^2$. Putting this together, one finds that 
the metastable vacuum $\phi=0$ has survived the high temperatures 
of the early Universe if
\begin{equation}
S_3/T\ga 230.
\end{equation}
For a given choice of masses $m$ and $M_3$, we have computed the 
minimum of $S_3/T$ as a function of temperature.

The resulting minimum
values are shown as contours in the  $(m/\sqrt{3},M_3)$--plane in
Fig.~4. The interpretation of these results is the following: in the
regions where for $T=0$ the action $S_4$ was very large, corresponding
to very large values of $\phi(0)$ in the bounce solution, also the
finite temperature bounce action is hoplessly large  due to the
exponentially large values of the bounce at the origin (since the
finite $T$ correction $\Delta V_T$ is positive, the value of $\phi(0)$
is even larger in the finite $T$ case). In the remaining 
region of small scalar masses, where the $T=0$ bounce had smaller
values of $\phi(0)$ ($\sim 10^3$--$10^5$~GeV), the following is found:
for $T\gg m$ and $M_3$, the barrier has a height $\propto \phi^2T^2$ for
$\phi<T$, while it behaves as $\bar
g_*\pi^2T^4/90+m^2(Q_\phi)\phi^2/2$ 
for $T>\phi$, and hence the potential can become negative only for
values $\phi\propto T^2/\sqrt{|m^2(Q_\phi)|}\gg T$. The width of the
bounce is now $\sim |m^2(Q_\phi)|^{-1/2}$, so that $S_3/T\propto
(T/\sqrt{|m^2(Q_\phi)|})^3$. This implies that at large temperatures,
it becomes harder to jump over the barrier. In fact, we usually find
that the optimum temperature for the transition is ${\cal O}(m$). As
can be seen in Fig.~4, this considerations imply that the lifetime of
the metastable vacuum  is again larger than the age of the Universe
with the exception of the small wedge corresponding to $m\ll M_3$ (to
the right of the contour labeled 2.4),
similarly as in the $T=0$ case.  

Notice that so far we have been assuming as an initial condition that
the scalar field $\phi$ is always sitting at the origin $\phi=0$ 
at high temperatures. Although this is the thermal equilibrium state
at large temperatures, 
it is however also possible    that  the scalar 
field $\phi$ is left initially far from the 
origin, {\it e.g.} at an early epoch near the end of inflation, and
may then roll towards the unbounded direction before reaching the
$\phi=0$ minimum. Due to the 
uncertainties connected with the hidden sector and  the form of 
the gravitational couplings at the Planck scale, it is difficult to 
determine the (model--dependent) form of the potential 
for $\phi$ during inflation \cite{gaillard},
 specially for large values of the field, although some mechanisms
which could adjust the initial $\phi$ values close to the origin have
been suggested \cite{dine}. 
In this 
paper we have then assumed that, when the temperature of the
Universe decreases to the point in which the unbounded direction
first appears in $V(\phi,T)$,
 the scalar field is already sitting close to its color conserving 
minimum\footnote{We thank E. Kolb for enlightening discussions 
about this point.}.

\vspace{0.5cm}
\begin{flushleft}
{\large\bf 5. Conclusions}
\end{flushleft}
\vspace{0.5cm}

In this paper we have investigated the constraints involving some 
combination of the sparticle masses at the weak scale arising from 
the requirement that the metastable and color conserving ground 
state $\phi=0$ is stable along flat directions where the effective 
potential $V(\phi)$ becomes unbounded from below for very large 
values of the scalar field. We have concentrated our attention in
 the direction in
field space analysed in ref.~\cite{olive}, and a similar analysis may
be applied to other dangerous directions considered in the literature 
\cite{color}. 

In spite of the presence of large regions of  the 
$(m/\sqrt{3},M_3)$--parameter space where the effective potential 
becomes unbounded 
unless some new physics  appears between the MSSM scale and the 
GUT scale \cite{olive}, we have pointed out  that such 
constraints are significantly weakened when considering 
the decay probability  for the metastable state $\phi=0$ 
along this unbounded direction. It turns out that the lifetime of this
state 
is longer than the
present age of the Universe and that it can survive both quantum 
tunneling, occuring at zero temperature, and the thermal excitations 
present in the early Universe. We then suggest  that no severe relations 
among  the
superpartner masses should necessarily be imposed to avoid disastrous color 
breaking: the metastable color conserving minimum is protected 
against jumping towards dangerous squark unbounded  directions 
and this makes the color breaking effects to be in practice not present.

\acknowledgments

It is a pleasure to thank A. Masiero for very useful discussions. AR was 
supported in part by the DOE and by NASA (NAG5-2788) at Fermilab. 

\newpage
{\large\bf Figure Captions}
\vspace{1 cm}

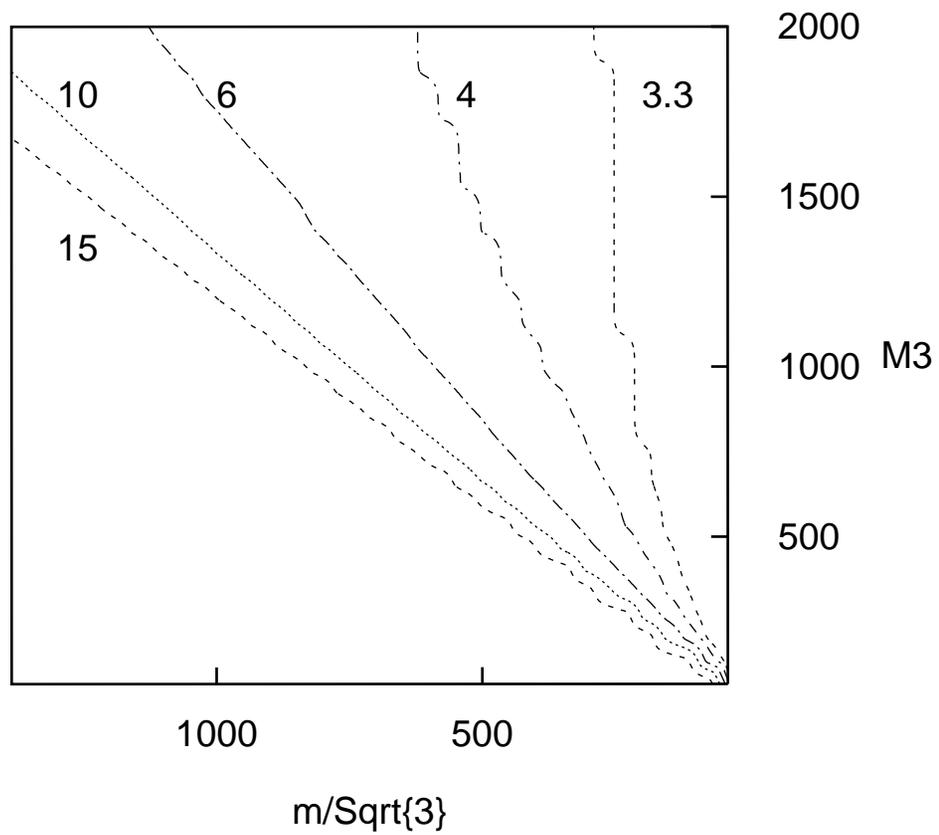
Figure 1:  Contours of $m^2(Q_0)=0$ for different choices of the scale 
$Q_0$, below which  new physics is then required, 
in the  $(m/\sqrt{3},M_3)$--plane.
\vspace{0.5cm}

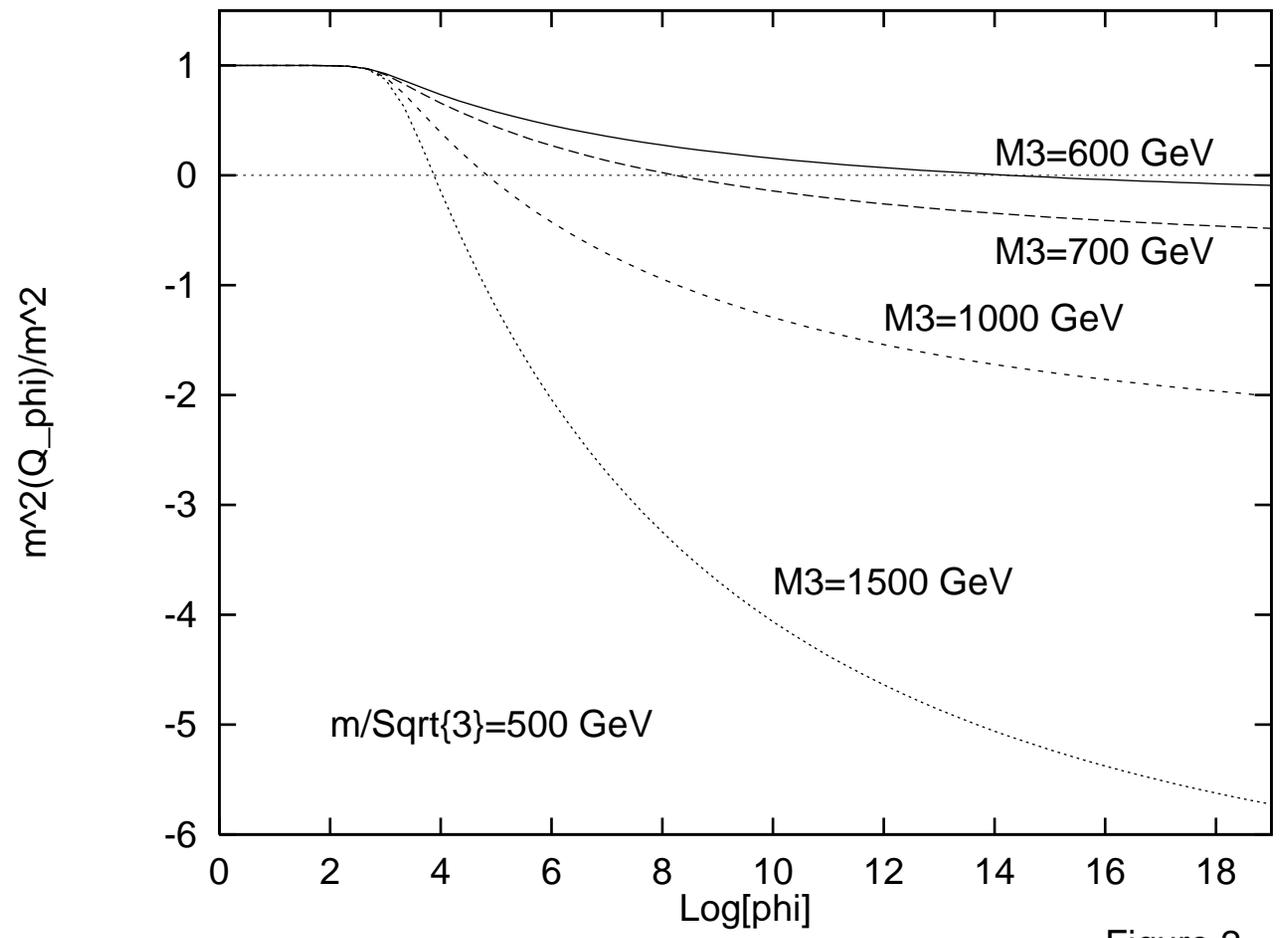
Figure 2: The plot of $m^2(Q_\phi)/m^2$ as a function of $\phi$
  for the particular 
choice of $m/\sqrt{3}=500$ 
GeV and $M_3=600$, 700, 1000, 1500~GeV. 
\vspace{0.5cm}

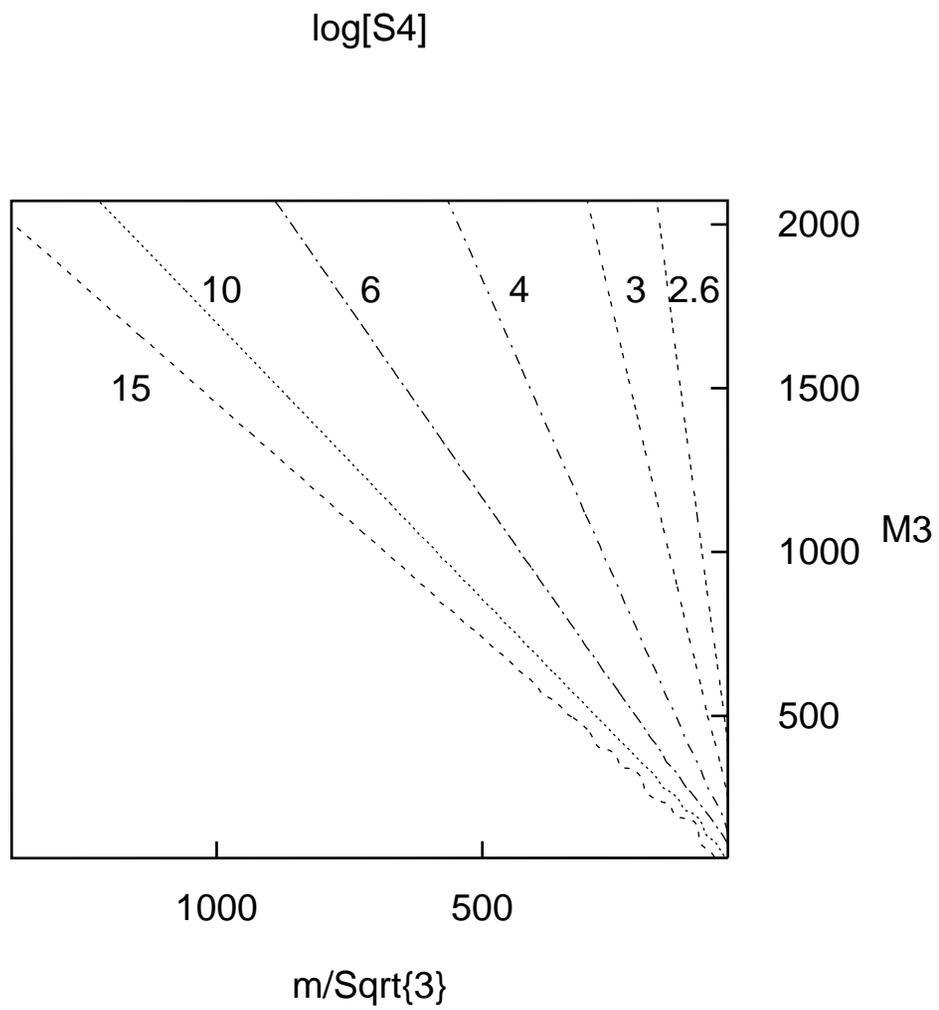
Figure 3: Contours  in the $(m/\sqrt{3},M_3)$--plane for different 
values of $\log_{10}(S_4)$, the critical value $S_4=400$ 
corresponding to the contour labelled by 2.6. 
\vspace{0.5cm}

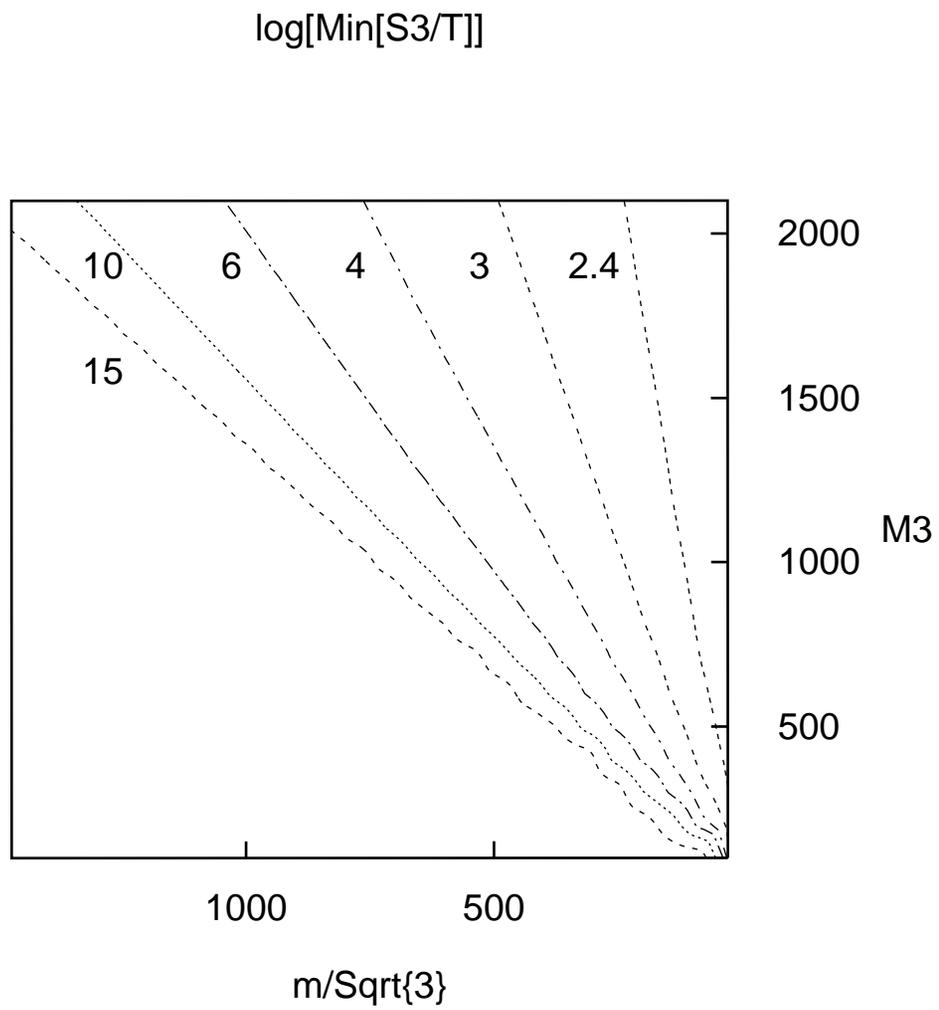
Figure 4: Contours of the minimum (as a function of the 
temperature) of $S_3/T$  in the $(m/\sqrt{3},M_3)$--plane. 
The critical value $S_3/T=230$ 
corresponding to the contour labelled by 2.4. 

\vspace{0.5cm}


\begin{references}

\begin{enumerate}

\bibitem{haber} For a review see, for instance, H.E. Haber and G.L. Kane, 
Phys. Rep. {\bf 117} (1985) 77. 

\bibitem{color}
C. Kounnas {\it et al.}, Nucl. Phys. {\bf B236} (1984) 438; 
J.M. Frere, D.R.T. Jones and S. Raby, Nucl. Phys. {\bf B222} (1983)
11; J.F. Gunion, H.E. Haber and M. Sher, Nucl. Phys. 
{\bf B306} (1988) 1; J.A. Casas, A. Lleyda and C. Mu\~noz, hep-ph/9507294. 

\bibitem{olive} T. Falk, K.A. Olive, L. Roszkowski and 
M. Srednicki, hep-ph/9510308. 

\bibitem{sher} See, for instance, M. Sher, Phys. Rep. {\bf 179} (1989) 274. 

\bibitem{meta} R. Flores and M. Sher, Phys. Rev. {\bf D27} (1982)
1679; 
M. Duncan, R. Philippe and M. Sher, Phys. Lett. 
{\bf B153} (1985) 165 and Phys. Lett. {\bf B209} (1988) 543(E); 
P. Arnold, Phys. Rev. {\bf D40} (1989) 613; G.W.  Anderson, 
Phys. Lett. {\bf B243} (1990) 265; P. Arnold and S. Vokos, 
Phys. Rev. {\bf D44} (1991) 3620; J.R. Espinosa and M. Quir\'os, 
Phys. Lett. {\bf B353} (1995) 257. 

\bibitem{gamb} G. Gamberini, G. Ridolfi and F. Zwirner, 
Nucl. Phys. {\bf B331} (1990) 331; B. de Carlos and J.A. Casas, 
Phys. Lett. {\bf B309} (1993) 320; 
J.A. Casas, J.R. Espinosa, M. Quir\'os and A. Riotto, 
Nucl. Phys. {\bf B436} (1995) 3. 

\bibitem{coleman} S. Coleman, Phys. Rev. {\bf D15} (1977) 2929; 
S. Coleman and C. Callan, Phys. Rev. {\bf D16} (1977) 1762. 

\bibitem{linde} A. Linde, phys. Lett. {\bf B70} (1977) 206; 
{\bf 100B} (1981) 37; A. Guth and E. Weinberg, Phys. Rev. 
{\bf D23} (1981) 876.

\bibitem{dj} L. Dolan and R. Jackiw, Phys. Rev. {\bf D9} (1974) 3320. 

\bibitem{gaillard} See, for instance, K. Gaillard, K.A. Olive and H. Murayama,
Phys. Lett. {\bf B355} (1995) 71.

\bibitem{dine} M. Dine, L. Randall and S. Thomas,
Phys. Rev. Lett. {\bf 75} (1995) 398.

\end{enumerate}
\end{references}
\end{document}